\documentclass%
[a4paper,pre,twocolumn,twoside,showpacs,showkeys,amsmath,amssymb,floatfix]%
{revtex4}

\pdfoutput=1
\usepackage{graphicx}
\usepackage{bm}
\usepackage{upgreek}
\usepackage{textcomp}
\usepackage{hyperref}

\begin{document}

\newcommand{\vect}[1]{\boldsymbol{#1}}
\newcommand{\tens}[1]{\mathbf{#1}}
\newcommand{\bnabla}{\boldsymbol\nabla}

\newcommand{\concv}{c_\text{v}}

\newcommand{\Ea}{E}
\newcommand{\Ha}{H}
\newcommand{\vectEa}{\vect E}
\newcommand{\vectHa}{\vect H}
\newcommand{\mue}{\mu_\textrm{e}}
\newcommand{\mui}{\mu_\textrm{i}}
\newcommand{\mud}{\mu_\textrm{d}}
\newcommand{\lambdae}{\lambda_\textrm{e}}
\newcommand{\lambdai}{\lambda_\textrm{i}}
\newcommand{\lambdad}{\lambda_\textrm{d}}
\newcommand{\epsilone}{\varepsilon_\textrm{e}}
\newcommand{\epsiloni}{\varepsilon_\textrm{i}}
\newcommand{\etae}{\eta_\textrm{e}}
\newcommand{\etai}{\eta_\textrm{i}}
\newcommand{\rhoe}{\rho_\textrm{e}}
\newcommand{\rhoi}{\rho_\textrm{i}}
\newcommand{\Rd}{a}

\newcommand{\phiE}{\theta_E}
\newcommand{\phiH}{\theta_H}

\newcommand{\resistance}{R}

\title{Anisotropy of magnetic emulsions induced by magnetic and electric fields}

\author{Yury I. Dikansky}
 \email{dikansky@mail.ru}
 \affiliation
 {Stavropol State University, Pushkin St., 1, Stavropol 355009, Russia}
\author{Alexander N. Tyatyushkin}
 \email{tan@imec.msu.ru}
 \thanks{the author to whom the correspondence should be addressed} 
 \affiliation
 {Institute of Mechanics, Moscow State University,
  Michurinskiy Ave., 1, Moscow 117192, Russia}
\author{Arthur R. Zakinyan}
 \email{zakinyan.a.r@mail.ru}
 \affiliation
 {Stavropol State University, Pushkin St., 1, Stavropol 355009, Russia}

\date{\today}

\begin{abstract}
The anisotropy of magnetic emulsions induced by simultaneously acting electric and magnetic fields is theoretically and experimentally investigated. Due to the anisotropy, the electric conductivity and magnetic permeability of a magnetic emulsion are no longer scalar coefficients, but are tensors. The electric conductivity and magnetic permeability tensors of sufficiently diluted emulsions in sufficiently weak electric and magnetic fields are found as functions of the electric and magnetic intensity vectors. The theoretically predicted induced anisotropy was verified experimentally. The experimental data are fitted in order to calculate the dimensions of the ellipsoidal drops of the emulsion distorted by the electric and magnetic fields. The dependence of the calculated fitting parameter on the intensity of the magnetic field is compared with the theoretical one. The results of the analysis of the experimental data are discussed.
\end{abstract}



\onecolumngrid
\vspace*{\textheight}%
\vspace*{2pc}{\hfill{\copyright}2011 American Physical Society}%
\vspace*{-4pc}%
\vspace*{-\textheight}%
\twocolumngrid

\maketitle

\section{Introduction}

\begin{figure}
\centerline{
\includegraphics[width=\columnwidth]{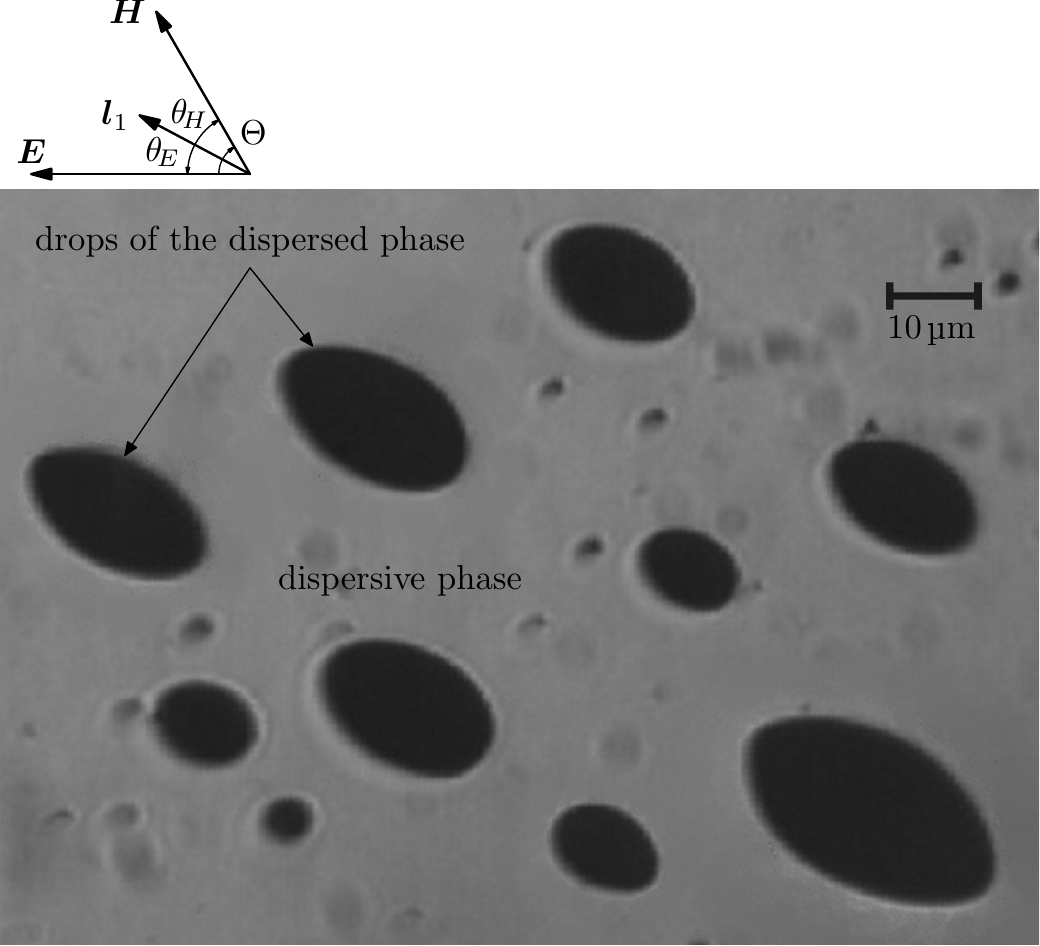}
}
\caption{Microphotograph of a magnetic emulsion affected simultaneously by electric and magnetic fields (the sketch above the microphotograph illustrates the orientations of the vectors defined in subsection II.A)}
\label{1}
\end{figure}

A magnetic emulsion is an emulsion at least one phase of which is a magnetic liquid (a close-up photograph of an emulsion,  dispersed phase of which is a magnetic liquid, is presented in Fig.~\ref{1}). 
Physical properties of magnetic emulsions have peculiarities that are not characteristic for homogeneous magnetic liquids. Drops of the dispersed phase of such emulsions form various structures under action of sufficiently strong magnetic fields \cite{Liu-etal}--\cite{Zhang&Widom}. Under action of magnetic fields with lower intensity magnitude, the drops deform, elongating along the intensity vector (see, e.g., \cite{Zakinyan-Dikansky}). Electric fields \cite{Dikanskii-etal}, hydrodynamic flows \cite{Sandre-etal}, and confined geometry \cite{Banerjee-etal} influence the deformation of the drops in magnetic fields. The scope of the present investigation is limited by simultaneous action of magnetic and electric fields that does not lead to formation of structures. 

The drops of the dispersed phase of a magnetic emulsion are spherical in the absence of electric and magnetic fields. Electric and magnetic fields distort the shape of the drops. The behavior of the drops in uniform constant or alternating electric and magnetic fields with parallel intensity vectors was investigated in \cite{Dikanskii-Tsebers-Shatskii}. The case of constant electric and magnetic fields with non-parallel intensity vectors was investigated in \cite{JCIS01}. Due to the distortion of the drops, magnetic emulsions become anisotropic in applied electric and magnetic fields. The goal of the present work is to investigate the  anisotropy of magnetic emulsions induced by simultaneously acting electric and magnetic fields.  

The case of diluted magnetic emulsions with weakly distorted or ellipsoidal drops is theoretically investigated in Sec.~\ref{II}. In Sec.~\ref{III}, the experimental investigation of the induced anisotropy is described. Analysis of the experimental data is provided in Sec.~\ref{IV}. The results of the investigation are summarized and discussed in Sec.~\ref{V}.

In what follows, normal italic letters denote scalar quantities, boldface italic letters denote vectors (the corresponding normal italic letters standing for their magnitudes), and upright boldface letters denote tensors of the second rank.

\section{Induced anisotropy}
\label{II}
\subsection{Single drop of the dispersed phase}

Consider a single drop of the dispersed phase of a magnetic emulsion in simultaneously acting uniform constant electric and magnetic fields. Under action of the electric and magnetic stresses, the drop always tends to stretch along the directions of the electric and magnetic intensity vectors, respectively. Aside from that, the electrohydrodynamic flow, which arises under action of the electric field, also deforms the drop so that the drop tends either to stretch or to shorten along the direction of the electric intensity vector. Let the radius of the drop in the absence of the fields be $\Rd$, the viscosity, electric conductivity, dielectric permittivity, and magnetic permeability of the dispersed and dispersive phases be respectively $\etai,\lambdai,\epsiloni,\mui$ and $\etae,\lambdae,\epsilone,\mue$, the surface tension of the interface between the phases be $\sigma$, and the intensity vectors of the magnetic and electric fields be respectively $\vectHa$ and $\vectEa$. If  $\lambdai$ and  $\lambdae$ are sufficiently small, then, to an approximation of the first order with respect to the dimensionless small parameters $\beta\Ha^2\Rd/\sigma$ and $\alpha\Ea^2\Rd/\sigma$, where
\begin{equation}
\alpha
=
\frac{9\lambdae}{8\pi}
\frac
 {16\etae+19\etai}
 {5\left(\etae+\etai\right)}
\frac
 {\epsiloni\lambdae-\lambdai\epsilone}
 {\left(2\lambdae+\lambdai\right)^2}
+
\frac{9\epsilone}{8\pi}
\left(
 \frac
  {\lambdae-\lambdai}
  {2\lambdae+\lambdai}
\right)^2
, 
\label{alpha}
\end{equation}
\begin{equation}
\beta
=
\frac{9\mue}{8\pi}
\left(
\frac{\mue-\mui}{2\mue+\mui}
\right)^2
,
\label{beta}
\end{equation}
the surface of the deformed drop is an ellipsoid, the semi-axes of which $a_1$, $a_2$, and $a_3$ are determined by the following expressions
\cite{JCIS01}:
\begin{align}
\dfrac{a_1^2-\Rd^2}{\Rd^2}
&=
\dfrac
 {\left(\alpha\Ea^2 + \beta\Ha^2 + 3{\cal{D}}\right)\Rd}
 {12\sigma}
,
\label{a}
\\*
\dfrac{a_2^2-\Rd^2}{\Rd^2}
&=
\dfrac
 {\left(\alpha\Ea^2 + \beta\Ha^2 - 3{\cal{D}}\right)\Rd}
 {12\sigma}
,
\label{b}
\\*
\dfrac{a_3^2-\Rd^2}{\Rd^2}
&=
-
\dfrac
 {\left(\alpha \Ea^2 + \beta \Ha^2\right)\Rd}
 {6\sigma}
.
\label{c}
\end{align}
The orientation of the ellipsoid is given by the unit vectors, $\vect{l}_1$, $\vect{l}_2$, and $\vect{l}_3$, directed along its main axes (see Fig.~\ref{1})
\begin{align}
\vect l_1
&=
\dfrac{\sin\phiH \; \Ha \vectEa - \sin\phiE \; \Ea \vectHa}
     {\left|\vectEa\times\vectHa\right|}
,
\label{l1}
\\*
\vect l_2
&=
\dfrac{-\cos\phiH \; \Ha \vectEa + \cos\phiE \; \Ea \vectHa}
     {\left|\vectEa\times\vectHa\right|}
,
\label{l2}
\\*
\vect l_3
&=
\dfrac{\vectEa\times\vectHa}{\left|\vectEa\times\vectHa\right|}
\label{l3}
,
\end{align}
where ${\cal{D}}$, $\phiH$, and $\phiE$ are determined as follows:
\begin{equation}
{\cal D}
=
\sqrt{
 \left(\alpha\Ea^2 - \beta\Ha^2\right)^2
 +
 4 \alpha \beta \left(\vectEa\cdot\vectHa\right)^2
}
,
\label{D}
\end{equation}
\begin{align}
\sin\left(2\phiH\right)
&=
\dfrac
 {2 \alpha \left|\vectEa\times\vectHa\right|\vectEa\cdot\vectHa}
 {\Ha^2 {\cal D}}
,
\label{psi_H}
\\*
\sin\left(2\phiE\right)
&=
-
\dfrac
 {2 \beta
  \left|\vectEa\times\vectHa\right|\vectEa\cdot\vectHa}
 {\Ea^2 {\cal D}}
\label{psi_E}
\end{align}
(note that $\phiH$ and $\phiE$ are oriented angles and $0\le\phiH\le\pi/2$ and $-\pi/2\le\phiE\le0$). Here, $\cdot$ and $\times$ denote the scalar and vector products, respectively, $|\vect{b}|$ denotes the magnitude of vector $\vect{b}$, and the formulas are written for the Gaussian system of units, which, unless otherwise stipulated, is used also in what follows. The conditions for the parameters of the problem $\Rd$, $\etai$, $\lambdai$, $\epsiloni$, $\mui$, $\etae$, $\lambdae$, $\epsilone$, $\mue$, $\sigma$, $\Ha$ and $\Ea$ that should be satisfied in order for Eqs.~(\ref{a})--(\ref{psi_E}) to be valid are written down in \cite{JCIS01}.

If $\vectEa$ and $\vectHa$ are parallel, 
\begin{align}
\frac{a_1^2-\Rd^2}{\Rd^2}
&=
\dfrac
 {\left(\alpha\Ea^2 + \beta\Ha^2\right)\Rd}
 {3\sigma}
,
\label{a-par}
\\*
\dfrac{a_2^2-\Rd^2}{\Rd^2}
&=
-
\dfrac
 {\left(\alpha\Ea^2 + \beta\Ha^2\right)\Rd}
 {6\sigma}
,
\label{b-par}
\\*
\dfrac{a_3^2-\Rd^2}{\Rd^2}
&=
-
\dfrac
 {\left(\alpha \Ea^2 + \beta \Ha^2\right)\Rd}
 {6\sigma}
,
\label{c-par}
\\
\vect l_1
&=
\dfrac{\vectEa}{\Ea}
=
\dfrac{\vectHa}{\Ha}
,
\label{l1-par}
\end{align}
and $\vect l_2$ and $\vect l_3$ are undetermined. This case corresponds to the degeneration into a prolate (if $\alpha>0$ or, when $\alpha\le0$, $|\alpha|\Ea^2<\beta\Ha^2$) or oblate (if $\alpha<0$ and $|\alpha|\Ea^2>\beta\Ha^2$) spheroid. If $\alpha<0$ and $|\alpha|\Ea^2=\beta\Ha^2$, the drops remains spherical in simultaneously acting constant uniform electric and magnetic fields with parallel intensities. The conditions for the drops to be either oblate or prolate or to remain spherical were found in \cite{Dikanskii-Tsebers-Shatskii}.

By introducing the dimensionless parameters $\Phi$ and $\mathcal{D}^*$ as follows
\begin{equation}
\Phi
=
\arctan\dfrac{\alpha\Ea^2}{\beta\Ha^2}
,
\label{Phi}
\end{equation}
\begin{equation}
\mathcal{D}^*
=
\dfrac{\mathcal{D}}{\sqrt{\alpha^2\Ea^4 + \beta^2\Ha^4}}
=
\sqrt{1+\sin(2\Phi)\cos(2\Theta)}
,
\label{D*}
\end{equation}
where $\Theta=\phiH-\phiE$ (see Fig.~\ref{1}), one obtains
\begin{align}
\dfrac{a_1^2}{\Rd^2}
&=
1
+
\dfrac{\Rd\sqrt{\alpha^2\Ea^4 + \beta^2\Ha^4}}{6\sigma}
\dfrac{\sin\Phi+\cos\Phi + 3\mathcal{D}^*}{2}
,
\label{a:}
\\*
\dfrac{a_2^2}{\Rd^2}
&=
1
+
\dfrac{\Rd\sqrt{\alpha^2\Ea^4 + \beta^2\Ha^4}}{6\sigma}
\dfrac{\sin\Phi+\cos\Phi - 3\mathcal{D}^*}{2}
,
\label{b:}
\\*
\dfrac{a_3^2}{\Rd^2}
&=
1
-
\dfrac{\Rd\sqrt{\alpha^2\Ea^4 + \beta^2\Ha^4}}{6\sigma}
\left(\sin\Phi+\cos\Phi\right)
,
\label{c:}
\end{align}
\begin{align}
\sin\left(2\phiH\right)
&=
\dfrac{2 \sin\Theta\cos\Theta\sin\Phi}{\mathcal{D}^*}
,
\label{psi_H:}
\\*
\sin\left(2\phiE\right)
&=
-
\dfrac{2 \sin\Theta\cos\Theta\cos\Phi}{\mathcal{D}^*}
.
\label{psi_E:}
\end{align}
The relations (\ref{a:})--(\ref{c:}) show that the smallness of the deformations is determined by the smallness of the only parameter $\Rd\sqrt{\alpha^2\Ea^4+\beta^2\Ha^4}/(6\sigma)$.

For small $x$, $1+x$ can be replaced by any equivalent (in the mathematical sense) expression, e.g., $1/(1-x)$, $1+\sin{x}$, $\cos\sqrt{x}$, or $\exp{x}$. Thus, for small parameter $\Rd\sqrt{\alpha^2\Ea^4+\beta^2\Ha^4}/(6\sigma)$, Eqs.~(\ref{a:})--(\ref{c:}) can be replaced by the following relations:
\begin{align}
\dfrac{a_1^2}{\Rd^2}
&=
\exp
 \left[
  \dfrac{\Rd\sqrt{\alpha^2\Ea^4 + \beta^2\Ha^4}}{6\sigma}
  \dfrac{\sin\Phi+\cos\Phi + 3\mathcal{D}^*}{2}
 \right]
,
\label{a:exp}
\\
\dfrac{a_2^2}{\Rd^2}
&=
\exp
 \left[
  \dfrac{\Rd\sqrt{\alpha^2\Ea^4 + \beta^2\Ha^4}}{6\sigma}
  \dfrac{\sin\Phi+\cos\Phi - 3\mathcal{D}^*}{2}
 \right]
,
\label{b:exp}
\\
\dfrac{a_3^2}{\Rd^2}
&=
\exp
 \left[
  -
  \dfrac{\Rd\sqrt{\alpha^2\Ea^4 + \beta^2\Ha^4}}{6\sigma}
  \left(\sin\Phi+\cos\Phi\right)
 \right]
.
\label{c:exp}
\end{align}
The advantage of the relations (\ref{a:exp})--(\ref{c:exp}) with respect to Eqs.~(\ref{a:})--(\ref{c:}) consists in the fact that, if the first are used, the product $a_1a_2a_3$, which is proportional to the volume of the drop, is exactly constant for arbitrary deformation (i.e., even outside the limits of validity of the relations), whereas, if the latter are used, the product $a_1a_2a_3$ is constant only with accuracy of the first order with respect to the small parameter $\Rd\sqrt{\alpha^2\Ea^4+\beta^2\Ha^4}/(6\sigma)$.

If the parameter $\Rd\sqrt{\alpha^2\Ea^4+\beta^2\Ha^4}/(6\sigma)$ is not small, the drop is strongly deformed and is no longer an ellipsoid. The orientation vectors, $\vect{l}_i$ ($n=1,2,3$), can also be defined for the case of arbitrary deformations if the deformed drop is orthotropic (i.e., has three orthogonal planes of symmetry). However, it is unknown whether the drop has such a symmetry if it is under simultaneous action of electric and magnetic fields. Note that Eqs.~(\ref{l1}) and (\ref{l2}) define $\phiE$ and $\phiH$ and are valid for an arbitrary orthotropic drop, whereas Eqs.~(\ref{psi_H}), (\ref{psi_E}), (\ref{psi_H:}), and (\ref{psi_E:}) determine $\phiE$ and $\phiH$ only for the case of small deformations of the drops.

\subsection{Electric conductivity and magnetic permeability}

Consider a dilute dispersion of identical ellipsoids with semi-axes $a_i$ of the same orientation given by the unit vectors $\vect{l}_i$ directed along the corresponding semi-axes ($i=1,2,3$). The tensors of electric conductivity and magnetic permeability, $\boldsymbol{\uplambda}$ and $\boldsymbol{\upmu}$, of such an anisotropic medium can be calculated with the help of method used in \cite{LL8} (see Sec.~9) for the calculation of the electric permittivity of a dilute suspension of spherical particles. Due to similarity of the equations (for static cases), this method can also be used for the calculation of the electric conductivity, magnetic permeability, and thermal conductivity of a dilute dispersion of spherical particles. Below, this method is generalized for the dilute dispersion of ellipsoids. The calculation of the electric conductivity of the dispersion is described in details and the magnetic permeability is calculated by analogy.

According to the above mentioned method, the conductivity tensor of the dispersion is defined by the relation
\begin{equation}
\dfrac{1}{V}\int\limits_V \vect{j} \,dV
=
\boldsymbol{\uplambda}\cdot\dfrac{1}{V}\int\limits_V \vect{E} \,dV
,
\label{averaging}
\end{equation}
where $V$ is the volume containing sufficiently large number of the ellipsoids, but sufficiently small with respect to the characteristic size of the problem, $\vect{j}$ and $\vect{E}$ are the electric current density and intensity within the phases of the dispersion, $\tens{S}\cdot\vect{b}$ denotes the contraction of tensor $\tens{S}$ with vector $\vect{b}$. Let $V_\textrm{e}$ and $V_\textrm{i}$ denote the parts of $V$ occupied by the dispersive and dispersed phases, respectively. Then,
\begin{multline}
\dfrac{1}{V}\int\limits_V \vect{j} \,dV
=
\dfrac{1}{V}\int\limits_{V_\textrm{e}} \lambdae\vect{E} \,dV
+
\dfrac{1}{V}\int\limits_{V_\textrm{i}} \lambdai\vect{E} \,dV
\\*
=
\lambdae\dfrac{1}{V}\int\limits_V \vect{E} \,dV
+
\left(\lambdai-\lambdae\right)
\dfrac{1}{V}\int\limits_{V_\textrm{i}} \vect{E} \,dV
.
\end{multline}
Since the dispersion is dilute, i.e., the ellipsoids are sufficiently distant from each other, each of them can be regarded as affected by the uniform field the intensity of which is equal to the average intensity of the emulsion in the vicinity of its position
\begin{equation*}
\dfrac{1}{V}\int\limits_V \vect{E} \,dV
.
\end{equation*}
Consequently, the electric field inside each ellipsoid can be regarded as uniform and  (cf. \cite{LL8}, Sec.~8) 
\begin{equation}
\int\limits_{V_\textrm{i}} \vect{E} \,dV
=
V_\textrm{i} 
\left(
 \tens{I}
 +
 \dfrac{\lambdai-\lambdae}{\lambdae}\tens{N}
\right)^{-1}
\cdot
\dfrac{1}{V}\int\limits_V \vect{E} \,dV
,
\end{equation}
where $\tens{I}$ is the unit tensor, $\tens{S}^{-1}$ denotes the inverse tensor with respect to the tensor $\tens{S}$, and $\tens{N}$ is the form-factor tensor of a single ellipsoid 
\begin{equation}
\tens{N} 
= 
N_{11}\vect{l}_1\vect{l}_1 
+
N_{22}\vect{l}_2\vect{l}_2 
+ 
N_{33}\vect{l}_3\vect{l}_3
,
\end{equation}
\begin{multline}
N_{ii}
=
\dfrac{a_1a_2a_3}{2} 
\int\limits_0^\infty
 \dfrac
  {d s}
  {\left(a_i^2+s\right)
   \sqrt{\left(a_1^2+s\right)\left(a_2^2+s\right)\left(a_3^2+s\right)}}
,
\\*
i=1,2,3
,
\end{multline}
where $\vect{b}\,\vect{d}$ denotes the dyadic product of vectors
$\vect{b}$ and $\vect{d}$. [The dimensionless quantities $N_{11}$, $N_{22}$, and $N_{33}$ are also referred to as demagnetization coefficients \cite{LL8} (see Sec.~8) or demagnetization factors \cite{Osborn}.] Thus, Eq.~(\ref{averaging}) takes the form
\begin{multline}
\left[
 \lambdae\tens{I}
 +
 \dfrac{V_\textrm{i}}{V}\left(\lambdai-\lambdae\right)
 \left(
  \tens{I}
  +
  \dfrac{\lambdai-\lambdae}{\lambdae}\tens{N}
 \right)^{-1}
\right]
\cdot
\int\limits_V \vect{E} \,dV
\\*
=
\boldsymbol{\uplambda}\cdot\int\limits_V \vect{E} \,dV
.
\end{multline}
Finally,
\begin{equation}
\boldsymbol{\uplambda}
=
\lambdae\tens{I}
+
\left(
 \dfrac{\tens{I}}{\lambdai-\lambdae}
 +
 \dfrac{\tens{N}}{\lambdae}
\right)^{-1}
\concv
,
\label{lambda}
\end{equation}
where $\concv=V_\textrm{i}/V$ is the volume concentration of the dispersed phase. In the same manner, one obtains
\begin{equation}
\boldsymbol{\upmu}
=
\mue\tens{I}
+
\left(
 \dfrac{\tens{I}}{\mui-\mue}
 +
 \dfrac{\tens{N}}{\mue}
\right)^{-1}
\concv
.
\label{mu}
\end{equation}
The tensors $\boldsymbol{\uplambda}$ and $\boldsymbol{\upmu}$ can be written in the form
\begin{gather}
\boldsymbol{\uplambda}
= 
\lambda_{11}\vect{l}_1\vect{l}_1 
+
\lambda_{22}\vect{l}_2\vect{l}_2 
+ 
\lambda_{33}\vect{l}_3\vect{l}_3
,
\label{lambda:g}
\\
\boldsymbol{\upmu}
= 
\mu_{11}\vect{l}_1\vect{l}_1 
+
\mu_{22}\vect{l}_2\vect{l}_2 
+ 
\mu_{33}\vect{l}_3\vect{l}_3
,
\label{mu:g}
\\
\lambda_{ii}
=
\lambdae
+
\left(
 \dfrac{1}{\lambdai-\lambdae}
 +
 \dfrac{N_{ii}}{\lambdae}
\right)^{-1}
\concv
,
\quad 
i=1,2,3
,
\label{lambdaii}
\\
\mu_{ii}
=
\mue
+
\left(
 \dfrac{1}{\mui-\mue}
 +
 \dfrac{N_{ii}}{\mue}
\right)^{-1}
\concv
,
\quad 
i=1,2,3
.
\label{muii}
\end{gather}

To an approximation of the first order with respect to the small parameters $\left(a_i^2-\Rd^2\right)/\Rd^2$ ($i=1,2,3$),
\begin{equation}
\tens{N}
=
\dfrac{1}{3} \tens{I}
-
\dfrac{1}{5}\dfrac{\tens{T}-\Rd^2\tens{I}}{\Rd^2}
,\quad
\tens{T} 
= 
a_1^2\vect{l}_1\vect{l}_1 
+
a_2^2\vect{l}_2\vect{l}_2 
+ 
a_3^2\vect{l}_3\vect{l}_3
.
\label{n}
\end{equation}
Here, it is used that, since the volume of the drop is constant, $a_1^2+a_2^2+a_3^2=3\Rd^2$ with accuracy of the first order with respect to the small parameters $\left(a_i^2-\Rd^2\right)/\Rd^2$. Substituting Eq.~(\ref{n}) into Eqs.~(\ref{lambda}) and (\ref{mu}) and using the smallness of the parameters 
$\left(a_i^2-\Rd^2\right)/\Rd^2$, one obtains
\begin{equation}
\boldsymbol{\uplambda}
=
\lambdad\tens{I}
+
\dfrac{9\lambdae\left(\lambdai-\lambdae\right)^2}
     {5\left(\lambdai+2\lambdae\right)^2}
\dfrac{\tens{T}-\Rd^2\tens{I}}{\Rd^2}
\concv
,
\label{lambda1}
\end{equation}
\begin{equation}
\boldsymbol{\upmu}
=
\mud\tens{I}
+
\dfrac{9\mue\left(\mui-\mue\right)^2}
     {5\left(\mui+2\mue\right)^2}
\dfrac{\tens{T}-\Rd^2\tens{I}}{\Rd^2}
\concv
,
\label{mu1}
\end{equation}
where
\begin{equation}
\lambdad
=
\lambdae
+
\dfrac{3\lambdae\left(\lambdai-\lambdae\right)}
     {\lambdai+2\lambdae}
\concv
,
\label{lambdad}
\end{equation}
\begin{equation}
\mud
=
\mue
+
\dfrac{3\mue\left(\mui-\mue\right)}
     {\mui+2\mue}
\concv
.
\label{mud}
\end{equation}

\subsection{Longitudinal and transversal components of the current density and magnetization}

Due to the anisotropy, the electric current density,
$\vect{j}=\boldsymbol{\uplambda}\cdot\vectEa$, and the
magnetic induction, $\vect{B}=\boldsymbol{\upmu}\cdot\vectHa$, in a magnetic emulsion have, in general, both longitudinal and transversal components relative to $\vectEa$ and $\vectHa$,
respectively. Using Eqs.~(\ref{lambda:g})--(\ref{muii}) and (\ref{l1})--(\ref{l3}), one obtains the following expressions for
the longitudinal and transversal components in the general case:
\begin{multline}
\vect{j}_{\!\parallel}
=
\lambdae\vectEa
+
\left(
 \dfrac{1}{\lambdai-\lambdae}
 +
 \dfrac{N_{11}}{\lambdae}
\right)^{-1}
\concv
\cos^2\phiE 
\vectEa
\\*
+
\left(
 \dfrac{1}{\lambdai-\lambdae}
 +
 \dfrac{N_{22}}{\lambdae}
\right)^{-1}
\concv
\sin^2\phiE 
\vectEa
,
\label{jpar:g}
\end{multline}
\begin{multline}
\vect{j}_\perp
=
\left[
 -
 \left(
  \dfrac{1}{\lambdai-\lambdae}
  +
  \dfrac{N_{11}}{\lambdae}
 \right)^{-1}
 +
 \left(
  \dfrac{1}{\lambdai-\lambdae}
  +
  \dfrac{N_{22}}{\lambdae}
 \right)^{-1}
\right]
\\*
\times
\concv
\sin\phiE\cos\phiE 
\dfrac
 {\left(\vectEa\times\vectHa\right)\times\vectEa}
 {\left|\vectEa\times\vectHa\right|}
,
\label{jper:g}
\end{multline}
\begin{multline}
\vect{B}_\parallel
=
\mue\vectHa
+
\left(
 \dfrac{1}{\mui-\mue}
 +
 \dfrac{N_{11}}{\mue}
\right)^{-1}
\concv
\cos^2\phiH
\vectHa
\\*
+
\left(
 \dfrac{1}{\mui-\mue}
 +
 \dfrac{N_{22}}{\mue}
\right)^{-1}
\concv
\sin^2\phiH
\vectHa
,
\label{Bpar:g}
\end{multline}
\begin{multline}
\vect{B}_\perp
=
\left[
 -
 \left(
  \dfrac{1}{\mui-\mue}
  +
  \dfrac{N_{11}}{\mue}
 \right)^{-1}
 +
 \left(
  \dfrac{1}{\mui-\mue}
  +
  \dfrac{N_{22}}{\mue}
 \right)^{-1}
\right]
\\*
\times
\concv
\sin\phiH\cos\phiH 
\dfrac
 {\left(\vectHa\times\vectEa\right)\times\vectHa}
 {\left|\vectEa\times\vectHa\right|}
.
\label{Mper:g}
\end{multline}
The longitudinal and transversal components for the case of small deformations of the drops are calculated with the use of Eqs.~(\ref{a})--(\ref{psi_E}) and (\ref{n})--(\ref{mu1}) and have the form
\begin{align}
\vect{j}_\parallel
= {}&
\lambdae\vectEa
+
\dfrac{3\lambdae\left(\lambdai-\lambdae\right)}
     {\lambdai+2\lambdae}
\concv\vectEa
\nonumber
\\*
&
+
\dfrac{9\lambdae\left(\lambdai-\lambdae\right)^2}
     {5\left(\lambdai+2\lambdae\right)^2}
\label{jpar}
\\*
&
\hphantom{{}+{}}\times
\dfrac
 {\left[
   2\alpha\Ea^4
   +3\beta\left(\vectEa\cdot\vectHa\right)^2
   -\beta\Ea^2\Ha^2
  \right] \Rd}
 {6\sigma\Ea^2}
\concv
\vectEa
\nonumber
,
\end{align}
\begin{equation}
\vect{j}_\perp
=
\dfrac{9\lambdae\left(\lambdai-\lambdae\right)^2}
     {5\left(\lambdai+2\lambdae\right)^2}
\dfrac
 {8\beta \Rd \vectEa\cdot\vectHa\
  \vectEa\times\left(\vectHa\times\vectEa\right)}
 {\sigma\Ea^2}
\concv
,
\label{jperp}
\end{equation}
\begin{align}
\vect{B}_\parallel
= {}&
\mue\vectHa
+
\dfrac{3\mue\left(\mui-\mue\right)}
     {4\pi\left(\mui+2\mue\right)}
\concv\vectHa
\nonumber
\\*
&
+
\dfrac{9\mue\left(\mui-\mue\right)^2}
     {5\left(\mui+2\mue\right)^2}
\label{Bpar}
\\*
&
\hphantom{{}+{}}\times
\dfrac
 {\left[
   2\beta\Ha^4
   +3\alpha\left(\vectEa\cdot\vectHa\right)^2
   -\alpha\Ea^2\Ha^2
  \right] \Rd}
 {6\pi\sigma\Ha^2}
\concv
\vectHa
\nonumber
,
\end{align}
\begin{equation}
\vect{B}_\perp
=
\dfrac{9\mue\left(\mui-\mue\right)^2}
     {5\left(\mui+2\mue\right)^2}
\dfrac
 {8\alpha \Rd \vectEa\cdot\vectHa\
  \vectHa\times\left(\vectEa\times\vectHa\right)}
 {\sigma\Ha^2}
\concv
.
\label{Bperp}
\end{equation}

It follows from Eqs.~(\ref{jperp}) and (\ref{Bperp}) that the transversal components are equal to zero when $\vectEa$ and $\vectHa$ are parallel or perpendicular and are maximal for given values of $\Ea$ and $\Ha$ when the angle between $\vectEa$ and $\vectHa$ is equal to $45^\circ$ or  $135^\circ$.

Consider a magnetic emulsion filling a rectangular parallelepiped. Let electric and magnetic fields in the parallelepiped be such that the electric and magnetic intensity vectors are parallel to some pair of opposite faces of the parallelepiped and are not parallel or perpendicular to each other, with the electric intensity vector being perpendicular to the second pair of opposite faces. Then, as it follows from Eq.~(\ref{jper:g}) or (\ref{jperp}), the electric current density should have a component perpendicular to the third pair of opposite faces, i.e., some electric current should flow from one of those faces to the other. This consideration hints an idea of an experimental verification of the induced anisotropy, which is described in the next section.

\section{Experiment}
\label{III}
\subsection{Experimental setup}

\begin{figure}[ht]
\centerline{
\includegraphics[width=\columnwidth]{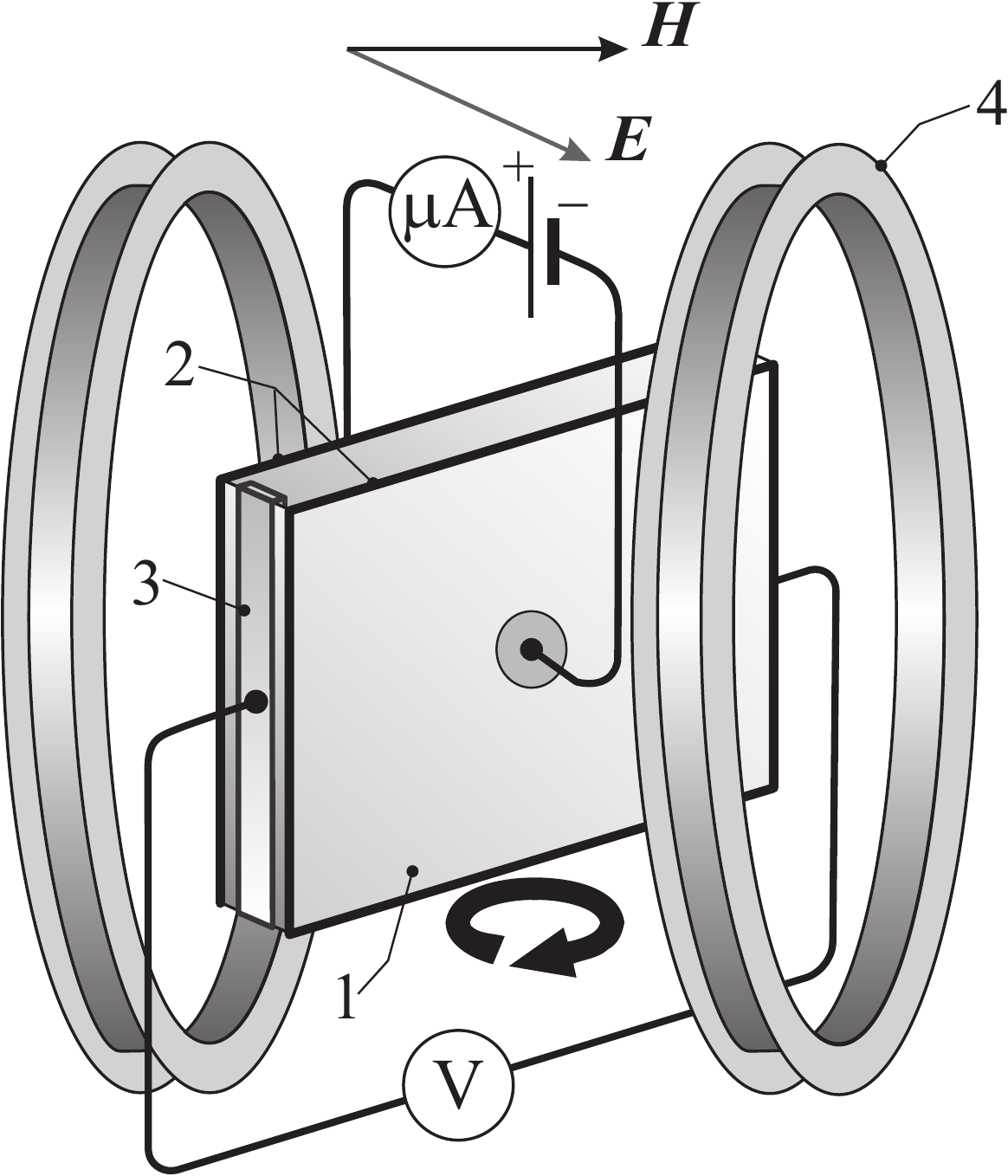}
}
\caption{Scheme of the experimental setup for the study of the anisotropy of electric conductivity}
\label{2}
\end{figure}

The experimental setup for investigation of induced anisotropy of  electric conductivity in magnetic emulsions is shown in Fig.~\ref{2}. The prepared emulsion is placed inside the cell 1. The interior of the cell is a rectangular parallelepiped with the dimensions 
${D_1}\times{D_2}\times{h}=33.3\times22.7\times1.2\;\text{mm}$. The internal surfaces of the faces 2 of the cell (the distance between which equals 1.2$\;$mm) have electrically conducting cover and serve as main electrodes. The lateral electrodes 3 that contact the emulsion from the side of the lateral faces serve to detect the transversal current with  measuring the voltage between them (transversal voltage). The Helmholtz coils 4 serve to create a magnetic field. The cell is able to rotate around a vertical axis, which allows varying and controlling the mutual orientation of the intensity vectors of the applied magnetic and electric fields with respect to each other.

The values of the parameters of the phases of the emulsion used in the experiments were $\mue=1$, $\mui=10.1$, $\lambdae=1.7\times10^{-9}\;\text{Sm/m}$, $\lambdai=8.1\times10^{-7}\;\text{Sm/m}$, $\epsilone=2.2$, $\epsiloni=5.8$ (all the quantities mentioned above measured with the bridge method), $\etae=0.0145\;\text{Pa}\cdot\text{s}$, $\etai=0.03\;\text{Pa}\cdot\text{s}$ (measured with a capillary viscometer), $\sigma=5\pm3\times10^{-6}\;\text{N/m}$ (measured with the method of form relaxation), $\concv=0.15$, $\Rd=6\pm2\times10^{-6}\;\text{m}$. These values correspond to $\alpha\approx0.8$ and $\beta\approx0.2$.

In order to verify that the possible diffusion of ions and contaminants between phases could be neglected, the conductivities were measured also after the emulsion was prepared and stratified into the fractions again. First, the conductivities of the components of the emulsion (oil and magnetic liquid) were measured. Then the prepared emulsion was affected by a constant electric field during an hour and was placed in non-uniform magnetic field, where it was stratified relatively soon. After that, the conductivities of the fractions were measured again. The conductivity of the magnetic liquid remained practically the same and that of the oil became higher by 4~\%. The increase in the conductivity of the oil was explained by the fact that the stratification was incomplete (some number of small magnetic liquid drops were detected with a microscope after the stratification). These additional measurements allowed neglecting the diffusion of ions and contaminants and using the values of the conductivities presented above in the calculations.

The electric field was created by a constant voltage $V=350\;\text{V}$ applied to the main electrodes, and thus its intensity was $E_\textrm{a}=2.92\times10^5\;\text{V/m}$. At this value of the electric field intensity, the strength of the electric current that passed through the magnetic emulsion was sufficient for reliable measurements, and, on the other hand, the electrohydrodynamic flows inside and near the drops of the dispersed phase did not cause an electrohydrodynamic flow in the entire emulsion. The electric current in the absence of magnetic filed, $I_{\parallel0}$, was equal to 15\;\textmu{A}. The intensity of the applied magnetic field in the cell, $H_\textrm{a}$, was equal to $1830\;\text{A/m}=23\;\text{Oe}$. As observations with an optical microscope showed, this value of the magnetic field intensity was the maximal one at which the coalescence of the drops and their transformation into threads and other structures could be neglected.

The transversal voltage (i.e., voltage between the lateral electrodes), $V_\perp$, and longitudinal current increment (i.e., current in the circuit of the main electrodes in the presence of the magnetic field minus that in absence of the magnetic field), $\Delta{I}_\parallel$, in the circuit of the main electrodes were simultaneously measured in three series of measurements with two digital multimeters GDM-8246 (one connected into the circuit of the main electrodes as an ammeter and the other, into the circuit of the lateral electrodes as a voltmeter).

\subsection{Estimation of the influence of the Hall effect}

Since the magnetic field acts on the moving ions that form current through the cell, the Hall effect takes place. This effect leads to the appearance of a voltage between upper and lower edge of the cell. However, if the variation of the current density within the cell caused by the Hall effect is not negligible, it can influence the measured transversal voltage as well. So it would be desirable to estimate the correction to the electric current density due to the Hall effect.

Without taking into account the Hall effect, the relation between the current density and the intensity of the electric field is as follows
\begin{equation}
\vect j
=
\sum_i z_i e n_i \vect v_i
=
\sum_i z_i e n_i m_i \vect E
,
\end{equation}
where $e=1.6021765\times10^{-19}\ \text{C}$ is the elementary positive charge, $z_i$, $n_i$, $\vect v_i$, and $m_i$ are the valency, number concentration, velocity, and mobility of the ion $i$, and the summation is performed over all the species of ions. If the correction to the current density due to the Hall effect is small, the relation between it and the induction of the magnetic field has the form
\newcommand{\vjH}{\vect{j}_\textrm{H}}
\begin{equation}
\vect{j}_\textrm{H}
=
\sum_i z_i e n_i m_i \dfrac{1}{c} \vect v_i \times \vect B
=
\sum_i z_i e n_i m_i^2 \dfrac{1}{c} \vect E \times \vect B
,
\end{equation}
where $c$ is the speed of light. Note that the corrections to the current density due to the Hall effect given by the positive and negative ions can completely compensate each other at some ratios of their mobilities and concentrations. In any case,
\begin{multline}
j_\textrm{H}
=
\left|\sum_i z_i e n_i m_i^2\right| 
\dfrac{1}{c} \left|\vect E \times \vect B\right|
\\
<
\left|\sum_i z_i e n_i m_i m_\textrm{max}\right| 
\dfrac{1}{c} E B
=
m_\textrm{max}\dfrac{1}{c} j B 
,
\end{multline}
\begin{equation}
\dfrac{j_\textrm{H}}{j}
<
m_\textrm{max}\dfrac{1}{c} B 
<
m_\textrm{max}\dfrac{1}{c} \mu_\textrm{max} H 
,
\end{equation}
where $m_\textrm{max}$ is the maximal absolute value of the mobility and $\mu_\textrm{max}$ is the maximal value of the magnetic permeability. Using the relation between the mobility of the ion $i$ and the viscosity of the liquid within which it moves
\begin{equation}
m_i
=
\dfrac{z_i e}{6\pi \eta r_i}
,
\end{equation}
where $r_i$ is the hydrodynamic radius of the ion $i$, one obtains the final formula for the estimation
\begin{equation}
\dfrac{j_\textrm{H}}{j}
<
\dfrac
 {z_\textrm{max} e}
 {6\pi \eta_\textrm{min} r_\textrm{min}} 
\dfrac{1}{c} \mu_\textrm{max} H
,
\label{Hall-estimate}
\end{equation}
where $z_\textrm{max}$ is the maximal absolute value of the valency and $\eta_\textrm{min}$ and $r_\textrm{min}$ are the minimal values of the viscosity and hydrodynamic radius. For SI, Eq.~(\ref{Hall-estimate}) is rewritten in the form
\begin{equation}
\dfrac{j_\textrm{H}}{j}
<
\dfrac
 {z_\textrm{max} e}
 {6\pi \eta_\textrm{min} r_\textrm{min}} 
\mu_\textrm{max} \mu_0 H 
,
\end{equation}
where $\mu_0=4\pi\times10^{-7}\ \text{V}\cdot\text{s}/(\text{A}\cdot\text{m})$ is the magnetic constant.

Taking $H=1830\ \text{A}/\text{m}$, $\mu_\textrm{max}=10.1$, $\eta_\textrm{min}=0.0145\ \text{Pa}\cdot\text{s}$, $r_\textrm{min}=10^{-10}\ \text{m}$, and $z_\textrm{max}=3$, one obtains $j_\textrm{H}/j<5\times10^{-10}$. Thus, the influence of the Hall effect in the experiments under consideration can be safely neglected.

\subsection{Experimental data}

\begin{figure}[htb]
\centerline{
\includegraphics[width=\columnwidth]{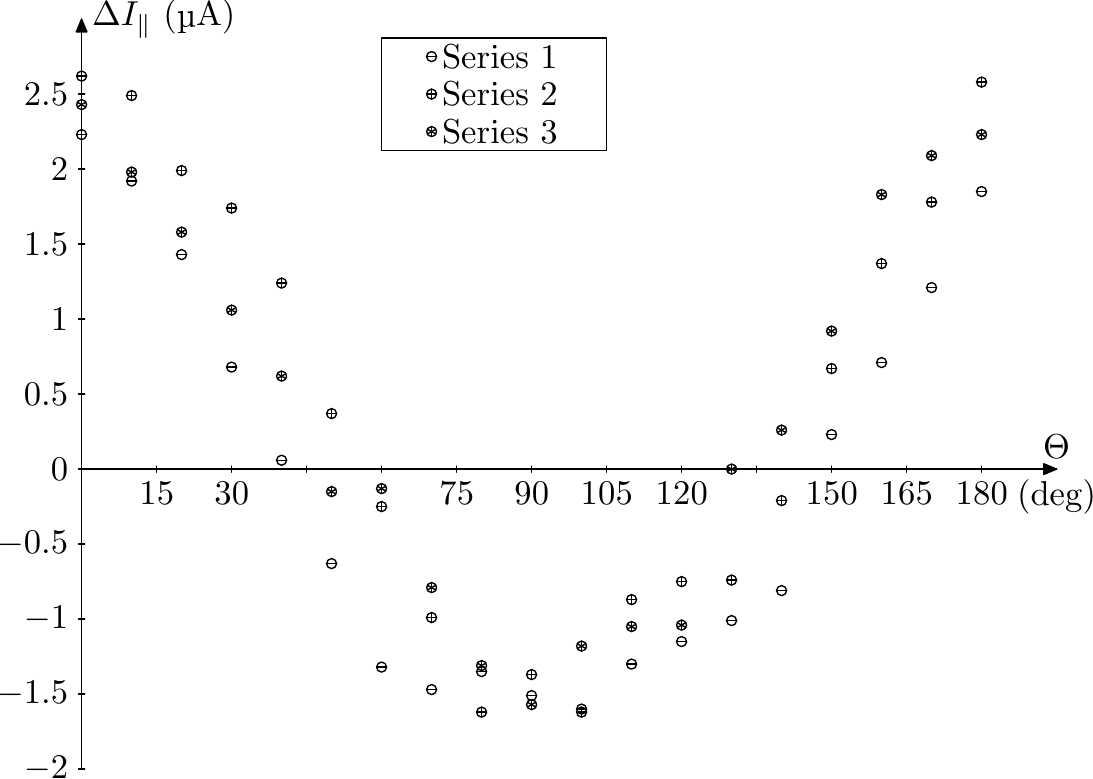}
}
\caption{Experimental dependence of the longitudinal current increment on the angle between the intensities of thew electric and magnetic fields}
\label{3}
\end{figure}
\begin{figure}[htb]
\centerline{
\includegraphics[width=\columnwidth]{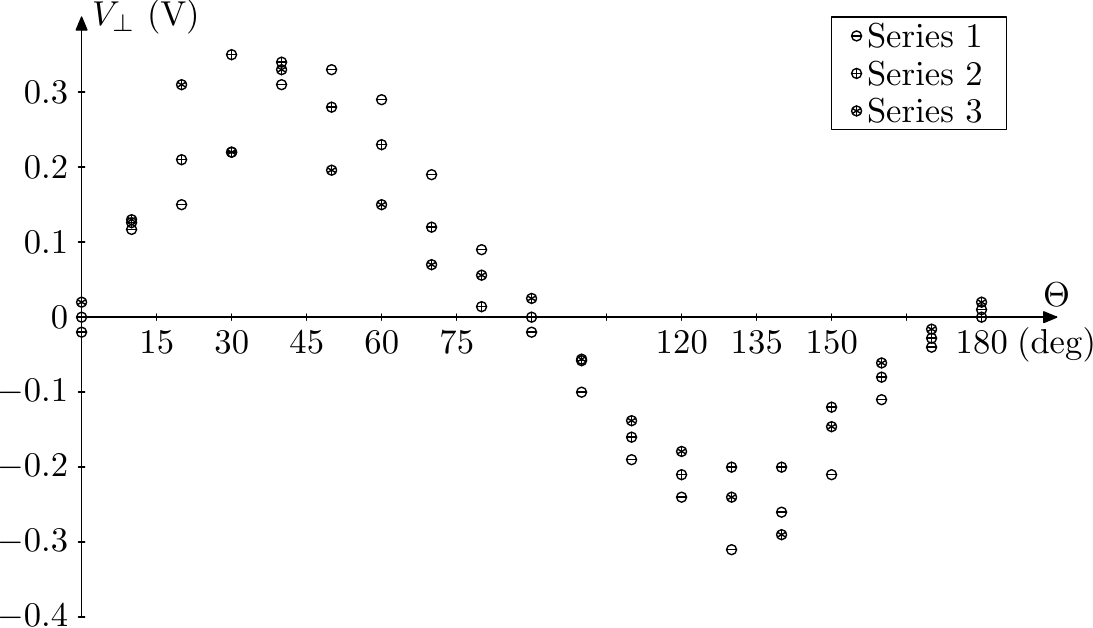}
}
\caption{Experimental dependence of the transversal voltage on the angle between the intensities of thew electric and magnetic fields}
\label{4}
\end{figure}
The transversal voltage and the longitudinal current increment are shown in Figs.~\ref{3}--\ref{4}.

Let us note that no transversal voltage was detected in similar experiments with pure magnetic liquids. The ability to become anisotropic under simultaneous action of non-parallel magnetic and electric fields is a characteristic peculiarity of magnetic emulsions. Thus, the experimental data confirm the theoretically predicted anisotropy of electric conductivity of magnetic emulsions caused by simultaneous action of magnetic and electric fields. 

Since the experimental verification of the induced anisotropy yielded numerical data, it would be worthy to analyze quantitatively at least some part of them. Such an analysis is presented in the next section.

\section{Analysis of the experimental data}
\label{IV}
\subsection{Electric and magnetic fields within the measuring cell}

Let $\resistance$ be the resistance of the voltmeter connected to the lateral electrodes. If the electric and magnetic fields were uniform, the transversal and longitudinal currents would be as follows:
\begin{equation}
I_\perp
=
V_\perp/\resistance
=
D_2 h j_\perp
,
\label{Iperp}
\end{equation}
\begin{equation}
I_\parallel
=
I_{\parallel0}
+
\Delta I_\parallel
=
D_1 D_2 j_\parallel
,
\label{Ipar}
\end{equation}
where $j_\perp$ and $j_\parallel$ would be determined by Eqs.~(\ref{jpar:g})--(\ref{jper:g}) or (\ref{jpar})--(\ref{jperp}). But the electric and magnetic fields inside the experimental cell are not uniform due to the edge effects (i.e., due to the the influence of the lateral electrodes and of the edges of the main electrodes). So Eqs.~(\ref{Iperp})--(\ref{Ipar}) with Eqs.~(\ref{jpar:g})--(\ref{jper:g}) or (\ref{jpar})--(\ref{jperp}) cannot be immediately used for calculation of $V_\perp$ and $\Delta{I}_\parallel$. Instead, a boundary value problem, which is formulated below, should be solved.

The electric and magnetic fields are determined by Maxwell's equations for constant electric and magnetic fields in a conducting magnetizable substance and by the material relations for the latter:
\begin{equation}
\bnabla\cdot\vect{j}
=
\boldsymbol{0}
,\quad
\bnabla\times\vect{E}
=
\boldsymbol{0}
,\quad
\vect{j}
=
\boldsymbol{\uplambda}\cdot\vect{E}
,
\label{eq:E}
\end{equation}
\begin{equation}
\bnabla\cdot\vect{B}
=
\boldsymbol{0}
,\quad
\bnabla\times\vect{H}
=
\boldsymbol{0}
,\quad
\vect{B}
=
\boldsymbol{\upmu}\cdot\vect{H}
.
\label{eq:H}
\end{equation}
Here, $\bnabla$ denotes nabla operator, $\vect{E}$ and $\vect{H}$ are the intensities of the electric and magnetic fields, $\vect{B}$ is the induction of the magnetic field,  $\vect{j}$ is the electric current density, and $\boldsymbol{\uplambda}$ and $\boldsymbol{\upmu}$ are determined by Eqs.~(\ref{lambda})--(\ref{mu}) or, if the drops of the dispersed phase are weakly distorted, by Eqs.~(\ref{n})--(\ref{mud}) and  (\ref{alpha})--(\ref{psi_E}).

The second equation in Eq.~(\ref{eq:E}) allows introducing the electric potential $\varphi$ determined by the relation $\vect{E}=-\bnabla\varphi$. Combining the latter with the first and third equations in Eq.~(\ref{eq:E}), one obtains the following equation for the electric potential $\varphi$:
\begin{equation}
\bnabla\cdot\left(\boldsymbol{\uplambda}\cdot\bnabla\varphi\right)
=
0
.
\label{eq:varphi}
\end{equation}
Note that Eq.~(\ref{eq:varphi}) is a non-linear equation since $\boldsymbol{\uplambda}$ depends on $\bnabla\varphi$. The boundary conditions for Eq.~(\ref{eq:varphi}) follow from the continuity of tangential component of the electric field intensity and of the normal component of the electric current density on the internal surfaces of the cell with regarding the conductivity of the electrodes as much greater than that of the emulsion:
\begin{align}
\varphi
=
\pm V/2
\quad
&\text{on the main electrodes}
,
\label{bc:phi:el1}
\\
\varphi
=
\left.\varphi\right|_{S_{\textrm{le}},S_{\textrm{re}}}
=
\text{const}
\quad
&\text{on the lateral electrodes}
, 
\label{bc:phi:el2}
\end{align}
\begin{equation}
\int\limits_{S_\textrm{le}} \vect{j} \cdot \vect{n} \,\mathrm{d}S
=
\int\limits_{S_\textrm{re}} \vect{j} \cdot \vect{n} \,\mathrm{d}S
=
\dfrac{V_\perp}{\resistance}
,
\end{equation}
and, with regarding the insulating parts of the internal surface of the cell as a perfect insulator: 
\begin{equation}
\vect{n}\cdot\boldsymbol{\uplambda}\cdot\bnabla\varphi
=
0
\quad
\text{on the insulating surfaces}
. 
\label{bc:phi:ins}
\end{equation}
Here, $\vect{n}$ is the unit normal vector to the corresponding surface, $S_{\textrm{le}}$ and $S_{\textrm{re}}$ denote the surfaces of the lateral electrodes that contact the magnetic emulsion.

By analogy to the electric field, one may introduce the scalar magnetic potential $\psi$ by the relation $\vect{H}=-\bnabla\psi$. Then the second equation in Eq.~(\ref{eq:H}) becomes identity, and the first and third equations yield the following non-linear equation for $\psi$:
\begin{equation}
\bnabla\cdot\left(\boldsymbol{\upmu}\cdot\bnabla\psi\right)
=
0
.
\label{eq:psi}
\end{equation}
The boundary conditions for Eq.~(\ref{eq:psi}) includes the conditions on the internal surfaces of the cell,
\begin{equation}
\left.\vect{n}\cdot\left(\bnabla\psi\right)\right|_\mathrm{n}
=
\left.
 \vect{n}\cdot\left(\boldsymbol{\upmu}\cdot\bnabla\psi\right)
\right|_\mathrm{m}
,
\label{bc:psi:n}
\end{equation}
\begin{equation}
\left.\vect{n}\times\left(\bnabla\psi\right)\right|_\mathrm{n}
=
\left.
 \vect{n}\times\left(\bnabla\psi\right)
\right|_\mathrm{m}
\label{bc:psi:t}
\end{equation}
(magnetization of any part of the cell except the emulsion is neglected), and the condition at infinity,
\begin{equation}
\psi(\vect{r})
\to
-
\vect{H}_\textrm{a}\cdot\vect{r}
\text{ as }
r\to\infty
\label{bc:psi:i}
\end{equation}
(the diameter of the Helmholtz coils is much greater than the size of the cell). Here, $\left.A\right|_\mathrm{m}$ and $\left.A\right|_\mathrm{n}$ denote the value of the quantity $A$ on the internal (directed to the magnetic emulsion) and external (directed to the non-magnetic wall of the cell) sides of the internal surfaces of the cell, $\vect{r}$ is the radius-vector with the origin at the center of the cell, $\vect{H}_\textrm{a}$ is the intensity of the uniform magnetic field in the cell created by the Helmholtz coils.

Solution of the system of equations (\ref{eq:varphi}) and (\ref{eq:psi}) with the boundary conditions (\ref{bc:phi:el1})--(\ref{bc:phi:ins}) and (\ref{bc:psi:n})--(\ref{bc:psi:i}) allows finding $V_\perp$ and $\Delta{I}_\parallel$: 
\begin{equation}
V_\perp
=
\left.\varphi\right|_{S_{\textrm{re}}}
-
\left.\varphi\right|_{S_{\textrm{le}}}
,
\end{equation}
\begin{equation}
\Delta{I}_\parallel
=
\int\limits_{S_\textrm{me}} \vect{j} \cdot \vect{n} \,\mathrm{d}S
-
I_{\parallel0}
,
\end{equation}
where $S_\textrm{me}$ denotes the surface of either of the two main electrodes that contacts the magnetic emulsion.

\subsection{Solution of the equations without taking into account the edge effects}

\begin{figure}
\centerline{
\includegraphics[width=\columnwidth]{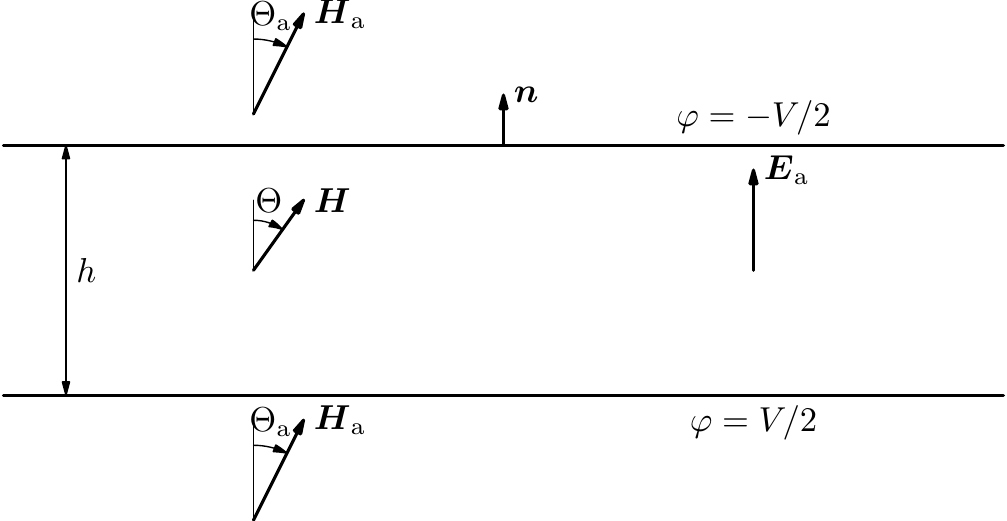}
}
\caption{Sketch of the orientations of the intensities of the electric and magnetic fields between and outside the infinite electrodes}
\label{5}
\end{figure}
Consider the case when the magnetic emulsion fills the layer of the thickness $h$ between two infinite plane parallel electrodes in an applied uniform magnetic field with a given intensity vector, $\vect{H}_\mathrm{a}$. Let the voltage between the electrodes be equal to $V$ (see Fig.~\ref{5}). Then, due to the symmetry, the intensity of the electric field, $\vect{E}$, and the electric current density, $\vect{j}$, inside the layer are as follows:
\begin{equation}
\vect{E}
=
\vect{E}_\mathrm{a}
=
\dfrac{V}{h} \vect{n}
,
\quad
\vect{j}
=
\dfrac{V}{h} \boldsymbol{\uplambda} \cdot \vect{n}
,
\end{equation}
where $\vect{n}$ is the internal normal to the positive electrode, and the intensities of the magnetic field inside and outside the layer are also uniform and determined by the boundary conditions  (\ref{bc:psi:n})--(\ref{bc:psi:i}). It follows from the boundary condition  (\ref{bc:psi:i}) that the magnetic field intensity outside the layer is equal to $\vect{H}_\mathrm{a}$. For the magnetic field intensity inside the layer, $\vect{H}$, the boundary conditions (\ref{bc:psi:n})--(\ref{bc:psi:t}) with taking into account Eqs.~(\ref{mu:g}), (\ref{muii}), and (\ref{l1})--(\ref{psi_E}) and the relation $\Theta=\phiH-\phiE$ yield the equations
\begin{equation}
H_\textrm{a} \sin \Theta_\textrm{a}
=
H \sin \Theta
,
\label{HTheta}
\end{equation}
\begin{multline}
H_\textrm{a} \cos \Theta_\textrm{a}
=
\dfrac{\mu_{11}+\mu_{22}}{2} H \cos \Theta
\\*
+
\dfrac{\mu_{11}-\mu_{22}}{2} H 
\left[\cos(2\phiE) \cos \Theta - \sin(2\phiE) \sin \Theta\right]
,
\end{multline}
from which it follows that
\begin{multline}
\cot \Theta_\textrm{a}
=
\dfrac{\mu_{11}+\mu_{22}}{2} \cot \Theta
\\*
+
\dfrac{\mu_{11}-\mu_{22}}{2}
\left[\cos(2\phiE) \cot \Theta - \sin(2\phiE)\right]
.
\label{Thetaa}
\end{multline}
Here, $\mu_{11}$ and $\mu_{22}$ are expressed in terms of $N_{11}$ and $N_{22}$ by Eq.~(\ref{muii}), $\Theta_\textrm{a}$ is the angle between $\vect{n}$ and $\vect{H}_\mathrm{a}$, and $\Theta$, between $\vect{n}$ and $\vect{H}$ (see Fig.~\ref{5}).

\subsection{Fitting of the dimensions of the distorted drops}

The major part of the longitudinal current flows through the cell sufficiently far from the edges of the main electrodes and from lateral electrodes. Thus, the influence of the edge effects may be neglected during its calculation if $h\ll\max(D_1,D_2)$. Whereas, in a calculation of the transversal current, the edge effects may not be neglected for any dimensions of the cell since all the transversal current goes through the regions near the lateral electrodes, where the electrical and magnetic fields are essentially nonuniform. So, only the data obtained in the measurements of the longitudinal current are used below.

For the treatment of the experimental data, Eqs.~(\ref{a:exp})--(\ref{c:exp}), (\ref{psi_H:}), and (\ref{psi_E:}) are used as fitting expressions for the semi-axes and orientation angles of the ellipsoidal drops. Let us note that this is a very strong assumption since those formulas are valid only for small deformations of the drops. By introducing the fitting parameter 
\begin{equation}
\tilde{a}
=
\exp
 \dfrac{\Rd\sqrt{\alpha^2\Ea^4 + \beta^2\Ha^4}}{12\sigma}
\label{tildea}
\end{equation}
and substituting Eq.~(\ref{D*}) into Eqs.~(\ref{a:})--(\ref{psi_E:}), one obtains
\begin{align}
\dfrac{a_1}{\Rd}
&=
\tilde{a}
 ^{
   \left.
    \left[
     \sin\Phi+\cos\Phi+3\sqrt{1+\sin(2\Phi)\cos(2\Theta)}
    \right]
   \right/2}
,
\label{a:f}
\\
\dfrac{a_2}{\Rd}
&=
\tilde{a}
 ^{
   \left.
    \left[
     \sin\Phi+\cos\Phi-3\sqrt{1+\sin(2\Phi)\cos(2\Theta)}
    \right]
   \right/2}
,
\label{b:f}
\\
\dfrac{a_3}{\Rd}
&=
a_3^*
=
\tilde{a}^{-\sin\Phi-\cos\Phi}
,
\label{c:f}
\end{align}
\begin{align}
\sin\left(2\phiH\right)
&=
\dfrac
 {\sin\Phi\sin(2\Theta)}
 {\sqrt{1+\sin(2\Phi)\cos(2\Theta)}}
,
\label{psi_H:f}
\\*
\sin\left(2\phiE\right)
&=
-
\dfrac
 {\cos\Phi\sin(2\Theta)}
 {\sqrt{1+\sin(2\Phi)\cos(2\Theta)}}
,
\label{psi_E:f}
\end{align}
where $\Phi$ is determined by Eq.~(\ref{Phi}). With taking into account that $\pi/2\le\phiE\le0$, it follows from (\ref{psi_E:f}) that 
\begin{equation}
\phiE
=
-
\dfrac{1}{2}
\arcsin
 \dfrac
  {\cos\Phi\sin(2\Theta)}
  {\sqrt{1+\sin(2\Phi)\cos(2\Theta)}}
.
\label{psi_E:ff}
\end{equation}
By eliminating $\tilde{a}$ from Eqs.~(\ref{a:f})--(\ref{c:f}), one obtains
\begin{align}
\dfrac{a_1}{\Rd}
&=
{a_{3}^*}
 ^{-
   \dfrac{1}{2}
   -
   \dfrac{3}{2}
   \left.
   \sqrt{1+\sin(2\Phi)\cos(2\Theta)}
   \right/
   (\sin\Phi+\cos\Phi)}
,
\label{a:f1}
\\
\dfrac{a_2}{\Rd}
&=
{a_{3}^*}
 ^{-
   \dfrac{1}{2}
   +
   \dfrac{3}{2}
   \left.
   \sqrt{1+\sin(2\Phi)\cos(2\Theta)}
   \right/
   (\sin\Phi+\cos\Phi)}
,
\label{b:f1}
\\
\dfrac{a_3}{\Rd}
&=
a_3^*
.
\label{c:f1}
\end{align}

With neglecting the edge effect, the longitudinal current through the cell can be expressed as follows 
\begin{multline}
I_\parallel
=
I_{\parallel0}+\Delta I_\parallel
\\*
=
\dfrac{V}{h} D_1 D_2 
\left(
 \dfrac{\lambda_{11}+\lambda_{22}}{2}
 +
  \dfrac{\lambda_{11}-\lambda_{22}}{2} \cos 2\phiE
\right)
,
\label{Ipar1}
\end{multline}
where $\lambda_{11}$ and $\lambda_{22}$ are expressed in terms of $N_{11}$ and $N_{22}$ by Eq.~(\ref{lambdaii}). According to \cite{Osborn},
\begin{align}
N_{11}
&=
\dfrac
 {\cos\phi\cos\vartheta}
 {\sin^3\vartheta k^2}
\left[F(\vartheta,k)-E(\vartheta,k)\right]
,
\label{N11}
\\
N_{22}
&=
\dfrac
 {\cos\phi\cos\vartheta}
 {\sin^3\vartheta k^2 (1-k^2)}
\left[
 E(\vartheta,k) 
 - 
 (1-k^2) F(\vartheta,k)
\vphantom{
\dfrac
 {k^2\sin^2\vartheta\cos\vartheta}
 {\cos\phi}
}
\right.
\nonumber
\\
&\hspace*{14em} 
\left.
- 
\dfrac
 {k^2\sin\vartheta\cos\vartheta}
 {\cos\phi}
\right]
,
\\
N_{33}
&=
\dfrac
 {\cos\phi\cos\vartheta}
 {\sin^3\vartheta(1-k^2)}
\left[
 \dfrac{\sin\vartheta\cos\phi}{\cos\vartheta}
 -
 E(\vartheta,k) 
\right]
,
\end{align}
where $F(\vartheta,k)$ and $E(\vartheta,k)$ are the incomplete elliptic integrals of the first and the second kind,
\begin{equation}
\cos\vartheta
=
\dfrac{a_3}{a_1}
,\quad
\cos\phi
=
\dfrac{a_2}{a_1}
,\quad
k
=
\dfrac{\sin\phi}{\sin\vartheta}
.
\label{k}
\end{equation}
\begin{figure}
\centerline{
\includegraphics[width=\columnwidth]{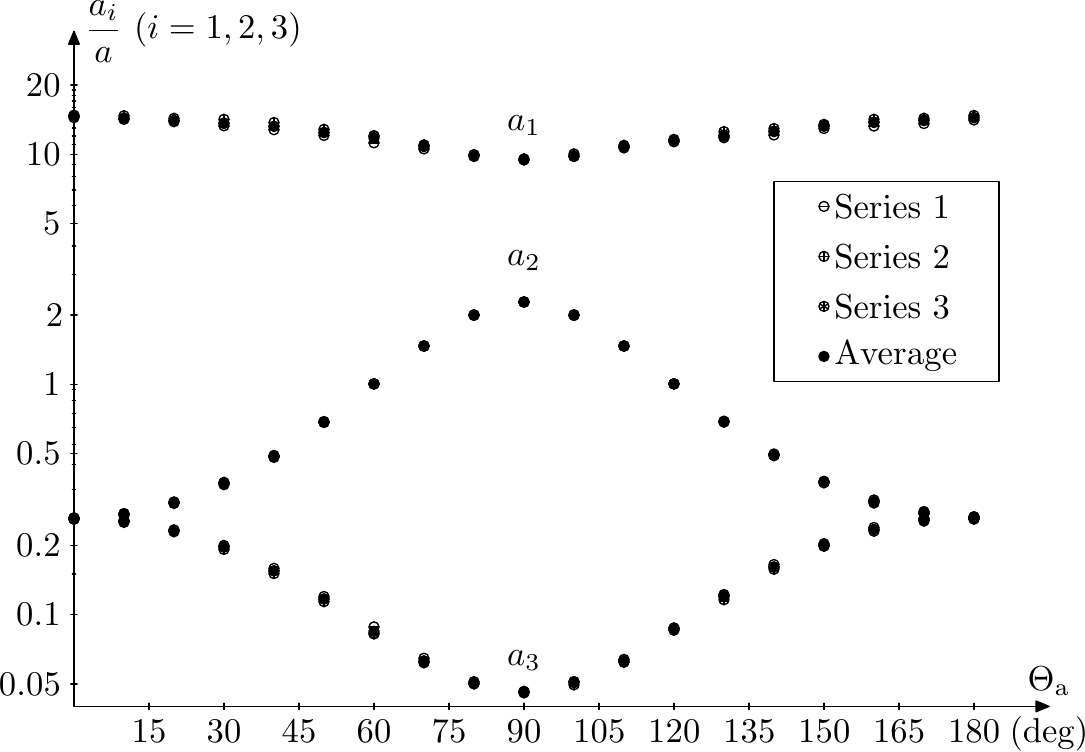}
}
\caption{Dependences of $a_1/\Rd$, $a_2/\Rd$, and $a_3/\Rd$ on $\Theta_\mathrm{a}$ calculated on the basis of the experimental data (semi-log plot)}
\label{6}
\end{figure}

\begin{figure}
\centerline{
\includegraphics[width=\columnwidth]{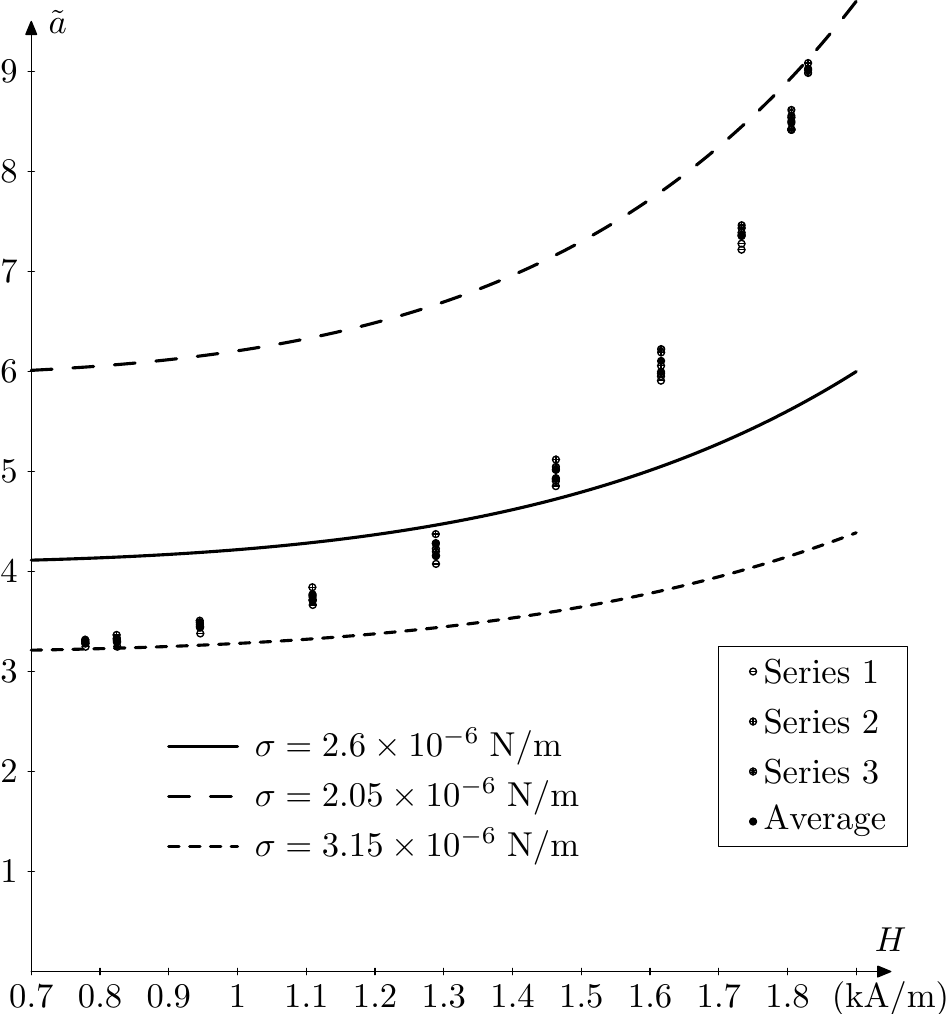}
}
\caption{Dependence of $\tilde{a}$ on $H$ (lines are the theoretical dependences for different values of the surface tension, circles are the dependencies calculated on the basis of the experimental data)}
\label{7}
\end{figure}
Substituting Eqs.~(\ref{lambdaii})--(\ref{muii}), (\ref{psi_E:ff})--(\ref{c:f1}), (\ref{Phi}), and (\ref{N11})--(\ref{k}) into Eqs.~(\ref{Thetaa}) and (\ref{Ipar1}), one obtains two equations of the form
\begin{align}
f_1(a_{3}^*,\Phi,\Theta)
&=
\cot\Theta_\textrm{a}
,
\label{Eq1}
\\*
f_2(a_{3}^*,\Phi,\Theta)
&=
I_\parallel
,
\label{Eq2}
\end{align}
where $f_1$ and $f_2$ are some functions of three variables. With the use of Eq.~(\ref{HTheta}), Eq.~(\ref{Phi}) can be rewritten in the form
\begin{equation}
\tan\Phi
=
\dfrac{\alpha E^2_\mathrm{a}}{\beta H^2_\mathrm{a}}
\dfrac{\sin^2\Theta}{\sin^2\Theta_\mathrm{a}}
.
\label{Eq3}
\end{equation}
Equations (\ref{Eq1})--(\ref{Eq3}) form a system of three equations with respect to the variables $a_{3}^*$, $\Phi$, and $\Theta$. Numerical solution of this system of equations for the measured values of $I_\parallel$ and $\Theta_\mathrm{a}$ allows obtaining the dependences of $a_1/\Rd$, $a_2/\Rd$, and $a_3/\Rd$ on $\Theta_\mathrm{a}$ (see Fig.~\ref{6}) and $\tilde{a}$ on $H$ (see Fig.~\ref{7}). (During the calculations, the corresponding equations were transformed from the Gaussian system of units to SI.)

Aside from the data obtained with fitting, the theoretical dependences calculated with the use of Eq.~(\ref{tildea}) are shown in Fig.~\ref{7}. For the calculations, the measured data indicated in Sec.~IIIA were used with taking $\Rd=6\times10^{-6}\;\text{m}$ and choosing two extreme values of $\sigma$ so that all the fitted data would fall between the extreme theoretical lines (dashed lines in Fig.~\ref{7}). Note that the  average and both extreme values of $\sigma$ fall within the measurement error interval of the measured value $\sigma=5\pm3\times10^{-6}\;\text{N/m}$.

\vspace{2pc}

\section{Summary and discussion}
\label{V}
Simultaneous action of constant magnetic and electric fields, which deform the drops of a magnetic emulsions, makes the emulsion anisotropic. Due to the anisotropy, the electric conductivity and magnetic permeability are no longer scalar quantities, but second rank tensors. For sufficiently dilute emulsions the dispersed phase of which can be regarded as constituted of identical ellipsoidal drops of the same orientation, the electric conductivity and magnetic permeability tensors are determined by Eqs.~(\ref{lambda:g}) and (\ref{mu:g}). If the drops are weakly distorted, the tensors are determined by Eqs.~(\ref{lambda1}) and (\ref{mu1}). 

If the intensity vectors of the electric and magnetic fields are not parallel or perpendicular, the transversal components of the electric current density and magnetic induction appear [see Eqs.~(\ref{jpar:g})--(\ref{Bperp})]. The latter property allows experimental verification of the theoretically predicted induced anisotropy. The experimental setup and the measurements are described in Sec.~III. The experimental data are presented in Figs.~\ref{3}--\ref{4}. Non-zero transversal voltage that takes place when the magnetic intensity vector is not collinear to the electric one (see Fig.~\ref{4}) is an unambiguous confirmation of the induced anisotropy.

Eqs.~(\ref{lambda1})--(\ref{mud}) together with Eqs.~(\ref{n}), (\ref{a:exp})--(\ref{c:exp}), (\ref{l1})--(\ref{l3}), (\ref{Phi})--(\ref{D*}), (\ref{psi_H:})--(\ref{psi_E:}), and (\ref{alpha})--(\ref{beta}) are exact theoretical relations that allow calculating the electric conductivity and magnetic permeability tensors for sufficiently dilute magnetic emulsions the drops of the dispersed phase of which are distorted weakly enough. However, in order to check quantitatively these theoretical relations in such emulsions, one should overcome considerable experimental difficulties. The results of measurements in magnetic emulsions are usually unstable since many undesirable processes take place in them, e.g., sedimentation, coagulation, and so on. Electric and magnetic fields make these processes more intensive. Due to these processes, measurements in magnetic emulsions lose reproducibility and some fluctuations of the measured values take place. So, in order to measure small variations in the current and to compare them with the exact theoretical relations, the fluctuations should be eliminated and the reproducibility should be restored. This could be reached by neutralization of the undesirable processes, which would lead to extremal complication of the experimental setup. The setup used in the experiments allowed detecting the induced anisotropy and performing reliable measurements in a rather concentrated emulsion with strongly distorted drops. So, in order to analyze the numerical data obtained in the experiments, the theoretical fitting formulas are deduced on the basis of very strong assumptions. The drops of the dispersed phase are regarded as ellipsoids whose sizes and orientations are determined by the relations valid only for weakly distorted drops. The emulsion is regarded as sufficiently dilute in order to neglect the electric, magnetic, and hydrodynamic interactions of its drops. The edge effects, due to which the electric and magnetic fields are not uniform, are neglected. The deduced fitting formulas allow determining the shape of the drops of the emulsion for various angles between electric and magnetic intensity vectors (see Fig.~\ref{6}). In spite of the above mentioned strong assumptions done during deducing the fitting formulas, the comparison of the theoretical dependence of the parameter $\tilde{a}$ on $H$ with that obtained with the use of the fitting (see Fig.~\ref{7}) demonstrates sufficiently good agreement: the dependences are both monotonously increasing and the theoretical and fitted values of the parameter $\tilde{a}$ coincide at least in the order of magnitude.

The results of the present investigation can also be used for estimation of the influence of the induced anisotropy in experiments with magnetic emulsions or in the design of devices in which magnetic emulsions are used.

\begin{acknowledgments}
This work is supported by Russian Foundation for Basic Research grants No.~\mbox{10-01-00015}, No.~\mbox{11-01-00051}, and No.~\mbox{10-02-90019-Bel}.
\end{acknowledgments}

\end{document}